\documentclass[12pt]{iopart}

\newtheorem{lemma}{Lemma}
\newtheorem{proposition}{Proposition}
\newtheorem{theorem}{Theorem}

\newcommand{\dif}{\mbox{\rm d}}

\newcommand{\ppsim}{\footnotesize\textcircled{\small$\wedge$}}
\def\Riem{{\rm Riem}}
\def\Ric{{\rm Ric}}
\def\r{{\rm r}}

\def\W{{\rm W}}
\newcommand{\be}{\begin{equation}}
\newcommand{\ee}{\end{equation}}

\begin{document}

\title[Characterization of spherically symmetric spacetimes] {An intrinsic
characterization of spherically symmetric spacetimes}

\author{Joan Josep Ferrando$^1$\
%\footnote[3]{To
%whom correspondence should be addressed (joan.ferrando@uv.es)}
and Juan Antonio S\'aez$^2$}

\address{$^1$\ Departament d'Astronomia i Astrof\'{\i}sica, Universitat
de Val\`encia, E-46100 Burjassot, Val\`encia, Spain.}

\address{$^2$\ Departament de Matem\`atiques per a l'Economia i l'Empresa,
Universitat de Val\`encia, E-46071 Val\`encia, Spain}

\ead{joan.ferrando@uv.es; juan.a.saez@uv.es}

\begin{abstract}
We give the necessary and sufficient (local) conditions for a metric
tensor to be a non conformally flat spherically symmetric solution.
These conditions exclusively involve explicit concomitants of the
Riemann tensor. As a direct application we obtain the {\em ideal}
labeling of the Schwarzschild, Reissner-Nordstr\"om and Lema\^itre-Tolman-Bondi
solutions.
\end{abstract}


%\submitto{\CQG}
%
\ \\

%\maketitle

\section{Introduction}
\label{sec-intro}

The spherically symmetric solutions of the Einstein equation have
played an essential role in further development of the General
Relativity theory. The simplicity of their geometry has allowed
analytic developments in modeling several astrophysical scenarios:
the Schwarzschild and Reissner-Nordstr\"om solutions as the
gravitational field exterior to a compact object, the geodesic
motion in the Schwarzschild geometry explaining the perihelion
advance and the bending of light, the relativistic equations of
stellar structure in spherically symmetric perfect fluid solutions,
the Friedmann-Lema\^itre-Robertson-Walker cosmological models, among
others (see, e.g., \cite{ple-kra}).

The spherically symmetric solutions continue being a subject of
study and, moreover, interest in them is increasing nowadays. For
instance, in numerical relativity they are an important tool in
testing fully general-relativistic time-dependent numerical codes
which evolve matter in strong gravitational fields (see, e.g.,
\cite{romero} an references therein). Nevertheless, despite the
extensive literature on spherically symmetric solutions, a fully
algorithmic characterization of these spacetimes has not been
accomplished. The goal of this paper is to attain such a
characterization.

The intrinsic characterization of spherically symmetric spacetimes
is an old problem established and partially solved by Takeno in 1952
\cite{takeno} (see also \cite{takeno-2}). In the abstract he
advanced: ``In this note a theory is developed concerning the
problem  of determining whether or not a space-time defined by a
line element arbitrarily given in any coordinate system is
spherically symmetric". And in the introduction Takeno adds: ``If
the answer of this problem is given by the tensor equations to be
satisfied by the curvature tensor specially when the equations
contain no tensor other than the metric tensor, the metric volume
element and the curvature tensor, we may say that the problem is
solved in the most desirable form". Finally, in his conclusions he
says:"...a theory concerning the discrimination of the spherically
symmetric spacetimes has been constructed. Although it is not of the
ideal form,...".

Given Takeno's work and his clear ideas two questions naturally
arise: Is it possible to solve the problem of characterizing the
spherically symmetric spacetimes in ``the ideal form"? Why doesn't
Takeno solve the problem in this ``most desirable form"?

The answer to the first question is affirmative as a consequence of
a result by Cartan \cite{cartan} who showed that each Riemannian
geometry may be characterized in terms of the Riemann tensor and its
covariant derivatives. Nevertheless, the labeling of metrics in the
``ideal form" starts with the beginning of the Riemannian geometry.
The characterization of locally flat Riemann spaces \cite{Riemann},
Riemann spaces with a maximal group of isometries \cite{Bianchi},
and locally conformally flat Riemann spaces \cite{Cotton}
\cite{Weyl} \cite{Schouten} was achieved before 1921. In all these
historic results, the conditions involve explicit concomitants of
the curvature tensor (Riemann, Weyl and Cotton tensors). Then, they
are {\em intrinsic} (depend solely on the metric tensor $g$) and,
besides, the {\em explicit} expression of these concomitants in
terms of $g$ is known. We say that a characterization with these two
properties, {\em intrinsic} and {\em explicit}, is an {\em ideal
characterization}.\footnote{The use of the appellation IDEAL (as an
acronym) is used by B. Coll and collaborators long ago. This acronym
seems to be adequate if the conditions involved, as well as being
Intrinsic and Explicit, are Deductive (no inference process is
necessary) and ALgorithmic (a flow chart with a finite number of
steps can be built).}

It is worth pointing out that an understanding of the algebraic
structure of the Weyl and Ricci tensors and the covariant
determination of their underlying geometry are necessary tools to
characterize spacetimes in an ideal form. But this comprehension of
the Riemann tensor had not been attained in the epoch of the
Takeno's aforementioned papers, which explains why he could not
achieve the characterization of the spherically symmetric spacetimes
in an ideal form.

The algebraic classification of the space-time symmetric tensors was
given by Churchill \cite{churchill}, and since the sixties a lot of
work has been devoted to studying the Ricci tensor from an algebraic
point of view (see, e.g., the paper by Pleba\'nski
\cite{plebanski}). However, the general covariant method to
determine the Ricci eigenvectors and their causal character was not
published until 1992 \cite{bcm}.

The pioneering papers by Petrov \cite{petrov} and Bel \cite{bel}
studied the algebraic classification of the Weyl tensor. Since then
several algorithms have been proposed to determine the Petrov-Bel
type. However, the general covariant expressions that give the
underlying geometry of the Weyl tensor for every Petrov-Bel type
were not published until 2001 \cite{fms}.

The accurate analysis of the underlying geometry of the Weyl and
Ricci tensors has allowed us to characterize in an ideal form
several families of spacetimes which are interesting from the
physical and/or geometrical point of view. We can quote the
characterizations of the Schwarzschild \cite{fsS},
Reissner-Nordstr\"{o}m \cite{fsD} and Kerr \cite{fsKerr} black
holes, among others (see references in \cite{fswarped}).

As Takeno claimed, the ideal characterization of spacetimes provides
an algorithmic way to test if a metric tensor, given in an arbitrary
coordinate system, is a specific solution of Einstein equations.
This can be useful to check wether a new solution is an already
known one. Also, it can be of interest in obtaining a fully
algorithmic characterization of the initial data which correspond to
such a solution. For instance, our ideal labeling of the
Schwarzschild geometry \cite{fsS} has allowed us to achieve the
algorithmic characterization of the Schwarzschild initial data
\cite{GP-VK-2}.

A spherically symmetric spacetime is a 2+2 warped product and this
quality determines the algebraic properties of the curvature tensor
of a metric with spherical symmetry. The approach that we present
here is based on this fact. We start from the ideal characterization
of the 2+2 warped spacetimes recently obtained \cite{fswarped}, and
we look for the complementary conditions on the curvature tensor
which lead to spherical symmetry. Section \ref{sec-warped} is
devoted to the study of the s-warped metrics, the class of the 2+2
warped spacetimes containing the spherically symmetric ones: we
present their ideal labeling and summarize some properties of their
curvature tensor.

In Section \ref{sec-ssst} we analyze the complementary constraints
that the spherical symmetry imposes and, when the Weyl tensor does
not vanish, we write these new conditions in terms of explicit
concomitant of the Weyl and Ricci tensors. Thus, we obtain our main
result: the ideal characterization of the non conformally flat
spherically symmetric spacetimes. Further, we give the algorithm to
determine the canonical elements of a spherically symmetric
spacetime from the Riemann tensor. An alternative characterization
which applies for a wide range of Ricci algebraic types is also
offered.

In Section \ref{sec-examples} we show that our result can be useful
to label specific spherically symmetric solutions in an ideal form.
As a consequence of our main theorem we recover the already known
characterizations of the Schwarzschild and Reissner-Nordstr\"{o}m
spacetimes \cite{fsS} \cite{fsD}, and we obtain the ideal labeling
of the Lema\^itre-Tolman-Bondi solution.

In this paper we work on an oriented spacetime with a metric tensor
$g$ of signature $\{-,+,+,+\}$. The Riemann, Weyl and Ricci tensors
and the scalar curvature are defined as given in \cite{kramer} and
denoted, respectively, by $\Riem(g)$, $W=\W(g)$, $R= \Ric(g)$ and
$\r$. For the metric product of two vectors we write $(x,y) =
g(x,y)$, and we put $x^2 = g(x,x)$. Other basic notation used in
this work is summarized in \ref{A-notation}.

\section{s-warped spacetimes}
\label{sec-warped}

A classification scheme for the warped product spacetimes was
proposed by Carot and da Costa \cite{Carot-costa} and the
isometry group that each class can admit was also studied. The energy tensor
algebraic types compatible with these classes were later analyzed
\cite{Haddow-carot}.

In a recent paper \cite{fswarped} we have given an invariant
classification of the spacetimes conformal to a 2+2 product and, for
the non conformally flat case, we have characterized each class in
an ideal form. The 2+2 warped metrics correspond to two of these
invariant classes, the {\em t-warped} and the {\em s-warped}
spacetimes.

In the s-warped spacetimes the warped plane is space-like, and the
metric tensor takes the expression \cite{fswarped}:
\be
g = v + e^{2 \lambda} \hat{h} \, , \qquad \hat{h}(\dif
\lambda)=0 \, ,
\ee
where $\lambda$ is the {\em warping factor} and where $v$ and
$\hat{h}$ denote two 2-dimensional metrics, Lorentzian and
Riemannian, respectively.

A s-warped spacetime has two principal planes, the time-like V
with projector $v$ and the space-like H with projector $h = e^{2
\lambda} \hat{h}$. The pair $(v,h)$ defines a  2+2 almost-product
structure with structure tensor $\Pi = v-h$. A 2+2 spacetime
structure is also determined by the {\it canonical} time-like
unitary 2-form $U$, volume element of the plane V, $U^2=v$.

\subsection{Invariant characterization of the s-warped spacetimes}
\label{subsec-invariant-warped}

The properties of the principal structure $(v,h)$ offer an invariant
characterization of the s-warped metrics \cite{fswarped}. More
precisely, the structure is umbilical and integrable, the plane V is
minimal, and the trace of the second fundamental form of the plane H
is a closed 1-form. These differential constraints can be explicitly
written in terms of either the canonical 2-form $U$ or the projector
$h$ and, consequently, one obtains an invariant labeling of the
s-warped metrics \cite{fswarped}:
\begin{proposition} \label{prop-s-warped-inv}
A {\rm 2+2} s-warped spacetime is characterized by one of the
following equivalent conditions:
\begin{enumerate}
\item There exists a simple, time-like and unitary {\rm 2}-form
$U$ ($U \! \cdot \! *U=0\, , \ \tr U^2 = 2$) such that
\begin{equation} \label{s-warped-invariant-U}
2 \, \nabla\! *\!U = - *\!U \ppsim b  \, , \qquad  \dif b=0\, ,
\qquad b \equiv U(\nabla \! \cdot \! U)\not=0 \, .
\end{equation}
\item There exists a space-like {\rm 2}-projector $h$ ($h^2 = h\, ,
\ \tr h = 2 \, , \ h(x,x) \geq 0 \, ,$ where $x$ in an arbitrary
time-like vector)  such that
\begin{equation} \label{s-warped-invariant-h}
2 \, \nabla h = h \stackrel{{\stackrel{23}{\sim}}}{\otimes} b  \,  ,
\qquad \dif b = 0  \,  , \qquad b \equiv  \nabla \! \cdot \! h \not = 0 \, .
\end{equation}
\end{enumerate}
Moreover, $U$ is the canonical {\rm 2}-form and $h$ is the projector
on the space-like plane. The warping factor $\lambda$ satisfies $2
\, \dif \lambda = -b$.
\end{proposition}

\subsection{Ricci and Weyl tensors}
\label{subsec-ricweyl-warped}

If we define $\hat{v} = e^{-2 \lambda}v$, we obtain that a s-warped
metric can be written as conformal to a 2+2 product metric, $g =
e^{2 \lambda} (\hat{v} + \hat{h})$, with  $\hat{h}(\dif \lambda)=0$.
Then, taking into account how the Weyl and Ricci
tensors change by a conformal transformation, we have
\cite{fswarped}:
\begin{proposition} \label{prop-Ricci-weyl-1}
Let $g =e^{2 \lambda } \, ( \hat{v} + \hat{h} )$ be a s-warped
metric and let $X$, $Y$ be the Gauss curvatures of $\hat{v} $ and
$\hat{h}$. Then,
\begin{enumerate}
\item The Weyl tensor is type D (or O) with double eigenvalue $\rho
\equiv - \frac{1}{6}e^{-2\lambda} (X+Y)$. In the type D case, the
canonical {\rm 2}-form $U$, $U^2 = v$, is the principal {\rm 2}-form
of the Weyl tensor which takes the expression:
\begin{equation} \label{Weyl-conf-pro}
W = \rho(3 S + G) \, ,  \quad S \equiv U \otimes U -
*U\otimes *U  \, , \quad G \equiv \frac12 \, g
\ppsim g\, .
\end{equation}
\item
The space-like principal plane is an eigen-plane of the Ricci tensor
$R$, that is, $R \cdot h = \mu \, h$ where $\mu$ is the associated
eigenvalue. The Ricci tensor takes the expression:
\begin{equation} \label{ricci-warped}
R = X \hat{v} + Y \hat{h} - 2 {\nabla} \dif{\lambda} - 2  \dif
\lambda \otimes \dif \lambda - \left[ {\Delta} \lambda - 2 {g}(\dif
\lambda, \dif \lambda) \right] \, {g} \, .
\end{equation}
\end{enumerate}
\end{proposition}

Note that the gradient of the warping factor is given by the vector
$b$ defined in proposition \ref{prop-s-warped-inv}, $2 \, \dif
\lambda = -b$. Then, if we take the trace in expression
(\ref{ricci-warped}) and we consider the expression of the Weyl
eigen-value given in point (i) of proposition
\ref{prop-Ricci-weyl-1}, we get:
\begin{equation} \label{constraint-2+2-conformal}
4 \, \rho + \frac{\rm r}{3}  = \nabla \cdot b + \frac12 \, b^2 \, ,
\end{equation}
where $\r$ is the scalar curvature, $\rho$ the double Weyl
eigenvalue and $b \equiv \nabla \cdot h$, $h$ being the projector on
the space-like plane.

On the other hand, we can obtain the Gauss curvature of the metric
$h$, $\kappa = e^{- 2 \lambda} Y$, as follows. Note that $h(b)=0$
and, consequently, $h^{\alpha \beta} \nabla_{\alpha} b_{\beta} = -
b^2$. Then, if we take the trace with the projector $h$ in
expression (\ref{ricci-warped}), we obtain:
\begin{equation} \label{gauss-1}
\kappa =  e^{- 2 \lambda} Y = \mu - \frac12  \nabla \cdot b \, ,
\end{equation}
where $\mu$ is the Ricci eigenvalue associated with the eigen-plane
H, $2 \mu = h^{\alpha \beta} R_{\alpha \beta}$. Moreover, if
we remove $\nabla \cdot b$ from equations
(\ref{constraint-2+2-conformal}) and (\ref{gauss-1}), we can
obtain an alternative expression for the Gauss curvature $\kappa$:
\begin{equation} \label{gauss-2}
\kappa = - 2 \, \rho- \frac{1}{6}\, \r + \mu + \frac14 \, b^2 \, .
\end{equation}

\subsection{Ideal characterization of the non conformally flat
s-warped metrics} \label{subsec-ideal-warped}

As a consequence of proposition \ref{prop-Ricci-weyl-1}, in the non
conformally flat case the Weyl tensor is type D and the canonical
2-form $U$ can be obtained from the Weyl tensor. Then, the invariant
characterization given in proposition \ref{prop-s-warped-inv} leads
to the following characterization of the s-warped metrics
\cite{fswarped}:
\begin{proposition} \label{prop-ideal-warped}
For a non conformally flat metric $g$ let $\rho$, $S$ and $\Phi$ be
the Weyl concomitants
\begin{equation} \label{warped-concomitants}
\rho \equiv - \left(\frac{1}{12} \tr W^3 \right)^{\frac{1}{3}} \neq
0  \, , \quad \, S \equiv \frac{1}{3 \rho} (W - \rho G) \, , \quad
\Phi \equiv \tr[S(\nabla \! \cdot \! S)] \, .
\end{equation}
Then, $g$ is a {\rm 2+2} s-warped metric if, and only if, it
satisfies:
\begin{eqnarray}
S^2 + S = 0 \, , \label{warped-typeD}\\[1mm]
2\, \nabla \! \cdot \! S + 3\, S(\nabla \! \cdot \! S) -
g \ppsim \Phi = 0 \, , \qquad \dif \Phi = 0 \, ,
\label{warped-conformal2+2} \\[1mm]
*S(\Phi; \Phi) =0, \qquad \ 2 \, S(\Phi, x; \Phi, x) -
\Phi^2 >  0 \, ,  \label{warped-s}
\end{eqnarray}
where $x$ is an arbitrary unitary time-like vector.
Moreover the warping factor
$\lambda$ satisfies $\Phi = - 2\, \dif \lambda$.
\end{proposition}
The notation used in this proposition is explained in detail in
\ref{A-notation}. Each of the conditions has a specific geometric
interpretation. The algebraic condition (\ref{warped-typeD}) means
that the Weyl tensor is of type D with real eigenvalues. The two
differential conditions (\ref{warped-conformal2+2}) ensure that the
metric is conformal to a 2+2 product metric with a principal
structure aligned with the Weyl tensor and, moreover, $\Phi = - 2\,
\dif \lambda$.  The differential conditions (\ref{warped-s}) mean
that $\Phi$ lies on the time-like principal plane or, equivalently,
that $h(\dif \lambda)=0$.

This proposition will be the starting point in
obtaining the characterization of the spherically symmetric
spacetimes.

\section{Characterizing spherically symmetric spacetimes}
\label{sec-ssst}

In a canonical coordinate system $(t,r,\theta , \varphi)$, the
metric line element of a spherically symmetric spacetime takes the
expression \cite{takeno-2} \cite{ple-kra}:
\begin{equation} \label{ss-can-1}
\dif s^2 = A \, \dif t^2 + 2\, C \, \dif t \, \dif r + B \, \dif r^2
+ e^{2 \lambda} \, \dif \Omega^2 \, ,
\end{equation}
where $\dif \Omega^2 = \dif \theta^2 + \sin^2 \, \dif \varphi^2$ is
the metric of the 2-sphere and $A$, $B$, $C$ and $\lambda$ are
smooth functions of $t$ and $r$.

In the coordinate expression (\ref{ss-can-1}) we can identify the
canonical form of a s-warped metric. More precisely, we have:
\begin{lemma} \label{lemma-can-ssst}
A spacetime is spherically symmetric if, and only if, the metric
tensor can be written as:
\begin{equation} \label{ss-can-2}
g = v + e^{2 \lambda} \sigma \, , \qquad \sigma(\dif \lambda)=0 \, ,
\end{equation}
where $v$ is an arbitrary {\rm 2}-dimensional Lorentzian metric, and
$\sigma$ is a {\rm 2}-dimensional Riemannian metric with a Gauss
curvature equal to one.
\end{lemma}

\subsection{Invariant characterization of the spherically symmetric
spacetimes} \label{subsec-invariant-ssst}

In the previous section we have characterized the s-warped
spacetimes. Now we must add new constraints in order to impose that
$\hat{h} = \sigma$, $\sigma$ being a metric with Gauss curvature
$Y=1$. As a consequence of (\ref{gauss-1}), this condition holds if,
and only, if, the Gauss curvature of $h$ is $\kappa = e^{-2
\lambda}>0$.

On the other hand, as $h(\dif \lambda)=0$, the condition $Y=1$
implies $h(\dif \kappa) =0$. And, conversely, as $Y$ is the
curvature of the metric $\hat{h}$ and $\kappa = e^{-2 \lambda}Y$,
conditions $h(\dif \kappa) =0$ and $\kappa>0$ imply that $Y$ is a
positive constant and, consequently, we can add an additive constant
to $\lambda$ in order to obtain a metric $\sigma$ with Gauss curvature $Y=1$.

Therefore, a s-product spacetime is spherically symmetric if, and
only if, the Gauss curvature $\kappa$ of the metric $h$ of the
space-like principal plane satisfies:
\begin{equation} \label{kappa-ssst-invar}
\kappa \equiv \kappa(h) \, > \, 0 \, , \qquad h (\dif \kappa) = 0 \, .
\end{equation}
Then, if we take into account proposition \ref{prop-s-warped-inv},
we obtain an invariant characterization of the metrics with
spherical symmetry:
\begin{proposition} \label{prop-ssst-inv}
A spherically symmetric spacetime is characterized by one of the
following equivalent conditions:
\begin{enumerate}
\item There exists a simple, time-like and unitary {\rm 2}-form $U$
which satisfies {\rm (\ref{s-warped-invariant-U})} and {\rm
(\ref{kappa-ssst-invar})}, where $\kappa(h)$ is the Gauss curvature
of the metric $h=U^2$.
\item There exists a space-like {\rm 2}-projector $h$ which satisfies
{\rm (\ref{s-warped-invariant-h})} and {\rm
(\ref{kappa-ssst-invar})}, where $\kappa(h)$ is the Gauss curvature
of $h$.
\end{enumerate}
Moreover, $U$ is the canonical {\rm 2}-form and $h$ is the projector
on the {\rm 2}-spheres. The warping factor $\lambda$ can be obtained
as $\lambda = - \frac12 \ln \kappa$, and $\sigma = \kappa \, h$ is
the metric of the {\rm 2}-sphere.
\end{proposition}

\subsection{Ideal characterization of the non conformally flat
spherically symmetric spacetimes} \label{subsec-ideal-ssst}

Proposition \ref{prop-ssst-inv} characterizes the spherically
symmetric spacetimes whatever their Riemann algebraic type. In order
to obtain an intrinsic labeling we must give either the canonical
2-form $U$ or the projector $h$ in terms the of the Weyl or Ricci
tensors. The use of the Ricci tensor requires the analysis of
several algebraic types and will be considered elsewhere
\cite{fs-ssst-Ricci}. Here we use the Weyl tensor.

When the spacetime is not conformally flat ($W\not=0$), the Weyl
tensor is of type D and, as stated in proposition
\ref{prop-Ricci-weyl-1}, the canonical 2-form $U$ is aligned with
the Weyl tensor. The explicit expression $U = U[W]$ of this Weyl
concomitant is given in \ref{A-deterU}. Then, as a consequence of
proposition \ref{prop-ssst-inv}, if we impose conditions
(\ref{s-warped-invariant-U}) and (\ref{kappa-ssst-invar}) to the
2-form $U = U[W]$ we get an intrinsic labeling of the non
conformally flat spherically symmetric spacetimes. Nevertheless, a
more accurate analysis of these constraints leads to more explicit
conditions in the Weyl tensor. Indeed, in \cite{fswarped} we have
shown that conditions (\ref{s-warped-invariant-U}) are equivalent to
equations (\ref{warped-typeD}), (\ref{warped-conformal2+2}) and
(\ref{warped-s}) which characterize the s-warped metrics in an ideal
form (see proposition \ref{prop-ideal-warped}). Thus, at this point
we have:
\begin{lemma} \label{lemma-ideal-invar}
A non conformally flat spacetime is spherically symmetric if, and
only if, the metric tensor satisfies the {\em intrinsic and
explicit} conditions {\rm (\ref{warped-typeD})}, {\rm
(\ref{warped-conformal2+2})} and {\rm (\ref{warped-s})}, and the
{\em invariant} conditions {\rm (\ref{kappa-ssst-invar})}.
\end{lemma}

Now we analyze the conditions (\ref{kappa-ssst-invar}) in order to
complete an ideal characterization of the spherical symmetry.

As commented after proposition \ref{prop-ideal-warped}, constraint
(\ref{warped-s}) states that $h(\Phi) = 0$. Moreover, under this
condition, the Weyl concomitant $\Phi$ defined in
(\ref{warped-concomitants}) equals the vector $b$ defined in
proposition \ref{prop-s-warped-inv}. Then, proposition
\ref{prop-ssst-inv} implies that for spherically symmetric
spacetimes the following identities hold:
\begin{equation}
\Phi = - 2\, \dif \lambda = b = \dif \ln \kappa \, .
\end{equation}
Note that the necessary conditions $h(\Phi)=0$ and $\Phi = \dif \ln
\kappa$ imply that $\dif \Phi = 0$ and  $h(\dif \kappa)=0$.
Consequently, we have:
\begin{lemma} \label{lemma-Phi-kappa}
In the set of constraints {\rm (\ref{warped-typeD})}, {\rm
(\ref{warped-conformal2+2})}, {\rm (\ref{warped-s})} and {\rm
(\ref{kappa-ssst-invar})} characterizing the spherically symmetric
spacetimes we can substitute conditions $\dif \Phi =0$ and $h(\dif
\kappa) =0$, by the condition $\Phi = \dif \ln \kappa$.
\end{lemma}

Finally, let us analyze expression (\ref{gauss-2}) of the curvature
$\kappa$. The scalars $\rho$ and $\r$, and the vector $b=\Phi$ are
explicit concomitants of the Weyl and Ricci tensors. How can the
Ricci eigenvalue $\mu$ be explicitly obtained? It satisfies $2 \mu =
h^{\alpha \beta} R_{\alpha \beta}$, where  the projector $h$ could
be obtained from the Weyl concomitant $U=U[W]$ given in lemma
\ref{lemma-determ-U-real} as $h = g - U^2$. Nevertheless, we try to
avoid the use of this expression of $U$ in our characterization
theorem because it involves an arbitrary 2-form $Z$ which could be
annoying in subsequent applications. Now, we get an alternative
expression for $\mu$ whatever the Ricci algebraic type by using a
mixed concomitant of the Ricci and Weyl tensors as follows.

The generic expression of a Ricci tensor which has H as an
eigen-plane with associated eigenvalue $\mu$ is:
\begin{equation}
R = A + \mu h \, ,
\end{equation}
where $A$ is a symmetric tensor on V, $A \cdot h =0$. Then, from the
expression (\ref{Weyl-conf-pro}) of $S$, we obtain:
\begin{equation}\label{be}
B \equiv R - S[R] - \frac12 \r g = \beta \, \Pi \, , \qquad \beta
\equiv \frac12 \r - 2 \mu \, .
\end{equation}
Thus, we have $\mu = \frac14 \r - \frac12 \beta$, where $\beta=0$
when $B=0$, and when $B\not=0$, $\beta$ can be computed as
\begin{equation} \label{beta}
\beta = \frac{\epsilon}{2} \sqrt{\tr B^2} , \qquad  \epsilon =
-\frac{B(x,x)}{|(B(x,x)| } \, ,
\end{equation}
$x$ being an arbitrary unitary time-like vector (note that the
structure tensor $\Pi = v-h$ satisfies $\Pi(x,x) < 0$ for every
time-like vector $x$).
This expression of $\beta$ allows us to obtain an explicit
expression of the Ricci eigenvalue $\mu$ and, substituting in
(\ref{gauss-2}), of the curvature $\kappa$.

All these considerations and lemma \ref{lemma-Phi-kappa} lead to the
following ideal characterization.

\begin{theorem} \label{theo-ideal-ssst}
Let $W =\W(g)$ and $R= \Ric(g)$ be the Weyl and Ricci tensors of a
non conformally flat metric $g$, and let $\rho$, $S$, $\Phi$, $\r$,
$B$ and $\kappa$ be the Riemann concomitants
\begin{eqnarray} \label{ssst-concomitants-1}
\hspace{-10mm} \rho \equiv - \left(\frac{1}{12} \tr W^3
\right)^{\frac{1}{3}}  \neq 0  \, , \quad \, S \equiv \frac{1}{3
\rho} (W - \rho G) \, , \quad \Phi
\equiv \tr[S(\nabla \! \cdot \! S)] \, , \\
\hspace{-10mm} \r \equiv \tr R \, , \quad B \equiv R - S[R] -
\frac12 \r g  \, , \quad  \kappa \equiv -2\, \rho + \frac{1}{12}\, r
+ \frac14 \, \Phi^2 + \frac{\epsilon}{4} \,  \sqrt{\tr B^2}  \, .
\label{ssst-concomitants-2}
\end{eqnarray}
where we put either $\epsilon =0$ if $B=0$ or $\epsilon=
-\frac{B(x,x)}{|B(x,x)|}$ if $B \neq 0$, $x$ being an arbitrary
unitary time-like vector.\\
Then, $g$ is spherically symmetric if, and only if,
it satisfies:
\begin{eqnarray}
S^2 + S = 0 \, , \label{ssst-typeD}\\[1mm]
2\, \nabla \! \cdot \! S + 3\, S(\nabla \! \cdot \! S) - g \ppsim
\Phi = 0 \, ,
\label{ssst-conformal2+2} \\[1mm]
*S(\Phi; \Phi) =0, \qquad \ 2 \, S(\Phi, x; \Phi, x) -
\Phi^2 \, >\,  0 \, ,  \label{ssst-warped-s} \\[1mm]
\kappa >0 \, , \qquad  \Phi = \dif \ln \kappa \, ,
\label{ssst-kappa+}
\end{eqnarray}
where $x$ is an arbitrary unitary time-like vector.
\end{theorem}

A metric tensor $g$ given in any coordinate system which satisfies
the conditions of theorem \ref{theo-ideal-ssst} is a spherically
symmetric solution and, as a consequence of lemma
\ref{lemma-can-ssst}, it admits the canonical form (\ref{ss-can-2}).
Can the canonical elements $v$, $\sigma$ and $\lambda$ be obtained
from the Riemann tensor?

As stated in the last sentence of proposition \ref{prop-ssst-inv},
the warping factor can be obtained from the Gauss curvature $\kappa$
as $\lambda = - \frac12 \ln \kappa$, and $\sigma = \kappa h$.
Moreover $h$ can be obtained as $h = g - v$, and $v= U^2$ where $U$
is the Weyl principal 2-form. Thus, we can state:
\begin{proposition} \label{prop-determin-canonical}
A non conformally flat metric $g$ with spherical symmetry admits the
canonical form:
\begin{equation} \label{ss-can-3}
g = v + \frac{1}{\kappa} \, \sigma \, , \qquad v = U^2 \, ,
\end{equation}
where $\kappa$ and $U$ are, respectively, the Riemann concomitants
defined in {\rm (\ref{ssst-concomitants-2})} and {\rm
(\ref{determ-U-real})}.
\end{proposition}

\subsection{On the other characterizations based more closely on the Ricci tensor}
\label{subsec-canonical}

Theorem \ref{theo-ideal-ssst} and proposition
\ref{prop-determin-canonical} apply when the Weyl tensor does not
vanish whatever the Ricci algebraic type. Under any algebraic
constraint of the Ricci tensor, alternative statements can be
deduced. Elsewhere \cite{fs-ssst-Ricci} we analyze in detail other
characterization theorems based more closely on the Ricci tensor.
Here, we illustrate these approaches by imposing an algebraic
condition which is valid for a wide range of Ricci types and which
avoids the use of the canonical 2-form $U$ in obtaining the
canonical elements.

Let us suppose that the 2-tensor $B$ defined in (\ref{be}) does not
vanish. Then $\beta \not=0$ and the structure tensor $\Pi$ can be
obtained as $\Pi = \frac{1}{\beta}B$. Thus, taking into account that
condition (\ref{ssst-warped-s}) is equivalent to $h(\Phi) =0$,
we can state:
\begin{proposition} \label{prop-ssst-alternative}
Let $g$ be a non conformally flat metric such that the Riemann
concomitant $B$ defined in {\rm (\ref{be})} does
not vanish. Then, the metric is spherically symmetric if, and only
if, it satisfies {\rm (\ref{ssst-typeD})}, {\rm
(\ref{ssst-conformal2+2})}, {\rm (\ref{ssst-kappa+})} and $
B(\Phi) = \beta \Phi \, , $
where $\beta$ is given in {\em (\ref{beta})}. \\
Then, the metric $g$ admits the canonical form {\rm
(\ref{ss-can-3})}, where $\kappa$ is defined in {\rm
(\ref{ssst-concomitants-2})} and $v$ is given by $v = \frac12(g +
\frac{1}{\beta} B) \, .$
\end{proposition}

\section{Labeling specific spherically symmetric solutions}
\label{sec-examples}

The ideal characterization given in theorem \ref{theo-ideal-ssst}
allows us to distinguish wether a non conformally flat metric given
in any coordinate system is spherically symmetric. Moreover, this
result could easily help us to obtain the ideal labeling of specific
spherically symmetric solutions. In this section we illustrate this
fact by recovering the already known characterizations of the
Schwarzschild \cite{fsS} and Reissner-Nordstr\"om \cite{fsD}
geometries, and by obtaining the ideal characterization of the
Lema\^itre-Tolman-Bondi solution.

In the two last cases, with a non vanishing Ricci tensor, an
alternative characterization based more closely on Ricci
concomitants could be acquired. Nevertheless, here we apply our
above theorem \ref{theo-ideal-ssst} and proposition
\ref{prop-ssst-alternative} based essentially on the Weyl tensor.
This way, the characterization theorem for the Reissner-Nordstr\"om
admits the Schwarzschild solution as a limiting case.

\subsection{Ideal characterization of the Schwarzschild solution}

The Jebsen-Birkhoff theorem \cite{jebsen} \cite{Birkhoff} states
that the Schwarzschild spacetime is the only spherically symmetric
vacuum solution. Consequently, we have:

\begin{lemma} \label{lemma-Schw}
A metric tensor $g$ is the Schwarzschild solution if, and only if, it
satisfies $R \equiv \Ric (g) = 0$ and the conditions of theorem {\rm
\ref{theo-ideal-ssst}}.
\end{lemma}

Let us see how the conditions in the above lemma can be simplified.
If we replace $S$ in terms of the Weyl tensor, condition
(\ref{ssst-conformal2+2}) is equivalent to
$$\rho \, \nabla \cdot W + W(\nabla \cdot W) -\frac13
g \ppsim \omega = 0 \, , \qquad \omega = \tr W(\nabla \cdot W) \, .$$
On the other hand, $\omega$ and $\Phi$ are related as $\Phi =
\frac{1}{9 \rho^2} \omega + \frac{2}{3}\, \dif \ln \rho.$

Then, under the vacuum condition, $R=0$, the Cotton tensor vanishes
and then $\nabla \cdot W =0$. Consequently,
(\ref{ssst-conformal2+2}) identically holds and $\Phi =
\frac{2}{3}\, \dif \ln \rho$. Then, the second condition in
(\ref{ssst-kappa+}) also holds and, moreover, the curvature $\kappa$
takes the expression $\kappa = \frac{1}{9}\, (\dif \ln \rho)^2 - 2
\rho$. Finally, we can substitute $\Phi$ by $\dif \rho$ in equation
(\ref{ssst-warped-s}). Thus, we recover the following ideal
characterization of the Schwarzschild geometry \cite{fsS}:
\begin{theorem} \label{theo-ideal-Sch}
Let $W =\W(g)$ and $R= \Ric(g)$ be the Weyl and Ricci tensors of a
non conformally flat metric $g$, and let $\rho$ and $S$ be the Weyl
concomitants
\begin{equation} \label{Sch-concomitants}
\rho \equiv - \left(\frac{1}{12} \tr W^3
\right)^{\frac{1}{3}}  , \qquad \, S \equiv \frac{1}{3
\rho} (W - \rho G) \, .
\end{equation}
Then, the necessary and sufficient conditions for $g$ to be the
Schwarzschild metric are:
\begin{eqnarray}
\Ric(g) =0 \, , \qquad \rho \not=0 \, , \qquad  S^2 + S = 0 \, ,
\label{Sch-1}\\[1mm]
\kappa \equiv \frac{1}{9}\, (\dif \ln \rho)^2 - 2
\rho >0 \, ,  \label{Sch-2} \\[1mm]
*S(\dif \rho; \dif \rho) =0, \qquad  2 \, S(\dif \rho , x; \dif \rho , x) -
(\dif \rho)^2 >  0 \, ,  \label{Sch-3}
\end{eqnarray}
where $x$ is an arbitrary unitary time-like vector.
\end{theorem}

\subsection{Ideal characterization of the Reissner-Nordstr\"om solution}

As a consequence of the generalized Birkhoff theorem (see
\cite{bona} and references therein), the only spherically symmetric
non null Einstein-Maxwell solution is the Reissner-Nordstr\"om
spacetime. The spherical symmetry implies that the electromagnetic
field is aligned with the Weyl tensor and moreover, as the principal
structure is integrable, the differential Rainich condition
identically holds \cite{rainich} \cite{fs-EM-align}. Then, we only
have to add the algebraic Rainich conditions. Consequently, we have:

\begin{lemma} \label{lemma-R-N}
A metric tensor $g$ is the Reissner-Nordstr\"om solution if, and only
if, it satisfies the conditions of theorem {\rm
\ref{theo-ideal-ssst}} and:
\begin{equation} \label{rainich}
\tr R = 0 \, , \qquad R^2 = \chi^2 g \, , \qquad R(x,x) \geq 0 \, ,
\end{equation}
where $x$ is an arbitrary unitary time-like vector and $\chi \equiv
-\frac12\sqrt{\tr R^2}$.
\end{lemma}

Again, the conditions in the above lemma can be simplified. As
commented above, the spherical symmetry implies the alignment of the
Ricci and Weyl tensors, an algebraic condition which is equivalent
to:
\begin{equation} \label{alignment}
S[R]+ R =0 \, .
\end{equation}

Under this constraint, the second Rainich condition in
(\ref{rainich}) is a consequence of the type D condition
(\ref{ssst-typeD}). Moreover, in \cite{fs-EM-align} we have shown
that the algebraic constraints (\ref{ssst-typeD}) and
(\ref{alignment}) imply the differential condition
(\ref{ssst-conformal2+2}) under the complementary algebraic
restriction $9 \rho^2 \not= \chi^2$.

On the other hand, for aligned non null Einstein-Maxwell solutions,
the Bianchi identities imply (when the Weyl tensor has real
eigenvalues) \cite{fs-EM-align}:
\begin{equation} \label{bianchi}
\dif \chi = 2\, \chi \Phi \, , \qquad 2\, \dif \rho = 3\, \rho \Phi
+ \chi \Pi(\Phi) \, .
\end{equation}
The second of these equations implies, under the algebraic condition
$9 \rho^2 \not= \chi^2$, that $h(\Phi)=0$ if, and only if, $h(\dif
\rho)=0$. Thus, we can substitute $\Phi$ by $\dif \rho$ in equation
(\ref{ssst-warped-s}). Moreover, in this case $\mu = - \chi$, and
from (\ref{bianchi}) we have $\Phi = \frac23 \dif \ln (\rho +
\chi)$, and we obtain that the curvature $\kappa$ given in
(\ref{ssst-concomitants-2}) becomes $\kappa = \frac19 \,(\dif \ln
\rho)^2 - 2\, \rho - \chi$. Thus, we recover the following ideal
characterization of the Reissner-Nordstr\"om geometry \cite{fsD}:
\begin{theorem} \label{theo-ideal-R-N}
Let $W =\W(g)$ and $R= \Ric(g)$ be the Weyl and Ricci tensors of a
non conformally flat metric $g$, and let $\rho$, $S$ and $\chi$ be
the Riemann concomitants
\begin{equation} \label{R-N-concomitants}
\hspace{-10mm} \rho \equiv - \left(\frac{1}{12} \tr W^3
\right)^{\frac{1}{3}}  \, , \quad \, S \equiv \frac{1}{3
\rho} (W - \rho G) \, , \quad \chi \equiv -\frac12\sqrt{\tr R^2}  \, .
\end{equation}
Then, the necessary and sufficient conditions for $g$ to be the
Reissner-Nordstr\"om metric are:
\begin{eqnarray}
\rho \not=0 \, , \quad  S^2 + S = 0 \, , \quad \tr R = 0  \, ,
\quad R(x,x) \geq 0 \, ,
 \label{R-N-1}\\[1mm]
S[R]+ R =0  \, , \qquad  9\, \rho^2 \not= \chi^2  \, ,
 \label{R-N-2}\\[1mm]
\kappa \equiv \frac19 \,(\dif \ln \rho)^2 - 2\, \rho - \chi >0 \, ,
\label{R-N-3} \\[1mm]
*S(\dif \rho; \dif \rho) =0, \qquad  2 \, S(\dif \rho , x;
\dif \rho , x) - (\dif \rho)^2 >  0 \, ,  \label{R-N-4}
\end{eqnarray}
where $x$ is an arbitrary unitary time-like vector.
\end{theorem}
If we make $\chi=0$ in the above theorem we recover the conditions
of theorem \ref{theo-ideal-Sch} which characterize the Schwarzschild
solution. Thus, theorem \ref{theo-ideal-R-N} includes the vacuum
case.

\subsection{Ideal characterization of the Lema\^itre-Tolman-Bondi
solution}

The Lema\^itre-Tolman-Bondi metrics (see, for example \cite{kramer})
are the spherically symmetric dust solutions such that the gradient
of the warped factor $\lambda$ is not collinear with the velocity of
the fluid. These constraints and the energy condition can be
explicitly written in terms of the Ricci tensor. Then we obtain:

\begin{lemma} \label{lemma-L-T-B}
A metric tensor $g$ is the Lema\^itre-Tolman-Bondi solution if, and only
if, it satisfies the conditions of theorem {\rm
\ref{theo-ideal-ssst}} and:
\begin{equation} \label{rainich-LTB}
R^2 = \frac14 \r^2 g \, , \qquad \r \equiv \tr R >0 \, , \qquad T
\ppsim \Phi \not= 0 \, ,
\end{equation}
where $T \equiv R -\frac12 \r g$.
\end{lemma}

We can simplify the conditions in this lemma considering the
alternative characterization given in proposition
\ref{prop-ssst-alternative} since now $B = - \frac{\r}{2} \Pi \not=
0$. Then $\beta =  - \frac{\r}{2}$, and the curvature $\kappa$ takes
the expression $\kappa = - 2\, \rho + \frac14 \Phi^2 - \frac16 \r$.
Thus, we obtain the following ideal characterization of the
Lema\^itre-Tolman-Bondi geometry:
\begin{theorem} \label{theo-ideal-LTB}
Let $W =\W(g)$ and $R= \Ric(g)$ be the Weyl and Ricci tensors of a
non conformally flat metric $g$, and let $\rho$, $S$, $\Phi$, $\r$,
$B$ and $\kappa$ be the Riemann concomitants
\begin{eqnarray} \label{LTB-concomitants-1}
\hspace{-10mm} \rho \equiv - \left(\frac{1}{12} \tr W^3
\right)^{\frac{1}{3}}  \neq 0  \, , \quad \, S \equiv \frac{1}{3
\rho} (W - \rho G) \, , \quad \Phi
\equiv \tr[S(\nabla \! \cdot \! S)] \, , \\
\hspace{-10mm} \r \equiv \tr R \, , \qquad B \equiv R - S[R] -
\frac12 \r g  \, , \qquad  T \equiv R -\frac12 \r g  \, .
\label{LTB-concomitants-2}
\end{eqnarray}
Then, the necessary and sufficient conditions for $g$ to be the
Lema\^itre-Tolman-Bondi metric are:
\begin{eqnarray}
\rho \not=0 \, , \quad  S^2 + S = 0 \, , \quad \r > 0  \, , \quad
R^2 = \frac14 \r^2 g \, , \quad T \ppsim \Phi \not= 0 \, ,
\label{LBT-1}\\[1mm]
2\, \nabla \! \cdot \! S + 3\, S(\nabla \! \cdot \! S) - g \ppsim
\Phi = 0  \, , \qquad  B(\Phi) + \frac{\r}{2} \Phi = 0 \, ,
\label{LBT-2} \\[1mm]
\kappa \equiv - 2\, \rho + \frac14 \Phi^2 - \frac16 \r >0 \, ,
\qquad   \Phi = \dif \ln \kappa \, .  \label{LBT-3}
\end{eqnarray}
\end{theorem}

\ack This work has been supported by the Spanish Ministerio de
Ciencia e Innovaci\'on, MICIN-FEDER project FIS2009-07705.

\appendix

\section{Notation}
\label{A-notation}

\begin{enumerate}
\item
{\bf Products and other formulas involving 2-tensors $A$ and $B$}:
\begin{enumerate}
\item
Composition as endomorphisms: $A \cdot B$,
\begin{equation}
(A \cdot B)^{\alpha}_{\ \beta} = A^{\alpha}_{\ \mu} B^{\mu}_{\
\beta}
\end{equation}
\item
Square and trace as an endomorphism:
\begin{equation}
A^2 = A \cdot A \, , \qquad \tr A = A^{\alpha}_{\ \alpha}.
\end{equation}
\item
Action on a vector $x$, as an endomorphism $A(x)$, and as a
quadratic form $ A(x,x)$:
\begin{equation}
A(x)^{\alpha} = A^{\alpha}_{\ \beta} x^{\beta}\, , \qquad A(x,x) =
A_{\alpha \beta} x^{\alpha} x^{\beta} \, .
\end{equation}
\item
Exterior product as double 1-forms: $A \ppsim B$ is the double
2-form,
\begin{equation}
(A \ppsim B)_{\alpha \beta \mu \nu} = A_{\alpha \mu} B_{\beta \nu} +
A_{\beta \nu} B_{\alpha \mu} - A_{\alpha \nu} B_{\beta \mu} -
A_{\beta \mu} B_{\alpha \nu} \, .
\end{equation}
\item
Exterior product with a vector $x$ as double 1-forms: $A \ppsim x$
is the vector-valued 2-form,
\begin{equation}
(A \ppsim x)_{\alpha,\, \mu \nu} = A_{\alpha \mu} x_{\nu} -
A_{\alpha \nu} x_{\mu}  \, .
\end{equation}
\item
Symmetrized tensorial product with a vector $x$ as double 1-forms:
$A \stackrel{{\stackrel{23}{\sim}}}{\otimes} x$ is the vector-valued
symmetric 2-tensor,
\begin{equation}
(A \stackrel{{\stackrel{23}{\sim}}}{\otimes} x)_{\alpha,\, \mu \nu}
= A_{\alpha \mu} x_{\nu} + A_{\alpha \nu} x_{\mu}  \, .
\end{equation}
\end{enumerate}

\item{\bf Products and other formulas involving double 2-forms $P$
and $Q$}:
\begin{enumerate}
\item
Composition as endomorphisms of the 2-forms space: $P \circ Q$,
\begin{equation}
(P \circ Q)^{\alpha \beta}_{\ \ \rho \sigma} = \frac12 P^{\alpha
\beta}_{\ \ \mu \nu} Q^{\mu \nu}_{\ \ \rho \sigma}
\end{equation}
\item
Square and trace as an endomorphism of the 2-forms space:
\begin{equation}
P^2 = P \circ P \, , \qquad \tr P = \frac12 P^{\alpha \beta}_{\ \
\alpha \beta}.
\end{equation}
\item
Action on a 2-form $X$, as an endomorphism $P(X)$, and as a
quadratic form $P(X,X)$,
\begin{equation}
P(X)_{\alpha \beta} = \frac12 P_{\alpha \beta}^{\ \ \mu \nu} X_{\mu
\nu} \, , \qquad P(X,X) = \frac14 P^{\alpha \beta \mu \nu} X_{\alpha
\beta} X_{\mu \nu}.
\end{equation}
\item
The Hodge dual operator is defined as the action of the, metric
volume element $\eta$ on a 2-form $F$ and a double 2-form $W$:
\begin{equation}
*F = \eta(F)  \, , \qquad  *W = \eta \circ W \, .
\end{equation}
\item
Action on two vectors $x$ and $y$, $P(x;y)$,
\begin{equation}
P(x;y)_{\alpha \beta} = P_{\alpha \mu \beta \nu} x^{\mu} y^{\nu} \,
.
\end{equation}
\item
Action on a vector-valued 2-form $Y$ as an endomorphism $P(Y)$,
\begin{equation}
P(Y)_{\lambda, \, \alpha \beta} = \frac12 P_{\alpha \beta}^{\ \ \mu
\nu} Y_{\lambda, \, \mu \nu} \, .
\end{equation}
\item
The trace of a vector-valued 2-form $Y$ is the 1-form $\tr Y$,
\begin{equation}
(\tr Y)_{\alpha} = g^{\lambda \mu} Y_{\lambda,\, \mu \alpha} \, .
\end{equation}
\item Action on a symmetric 2-tensor $B$ as an endomorphism $P[B]$,
\begin{equation}
P[B]_{\alpha \beta} = {{{P_{\alpha}}^{\mu}}_{\beta}}^{\nu} \ B_{\mu
\nu}
\end{equation}
\end{enumerate}
\end{enumerate}

\section{Covariant determination of the canonical 2-form $U$ of a type D
Weyl tensor with real eigenvalues} \label{A-deterU}

The explicit expression of the canonical 2-form $U$ of a type D Weyl
tensor using real formalism has been given in a recent paper
\cite{fsKerr}. Here we particularize it for the case of real
eigenvalues:
\begin{lemma} \label{lemma-determ-U-real}
For a Petrov-Bel type D Weyl tensor with real eigenvalue $\rho
\equiv - (\frac{1}{12} \tr W^3 )^{\frac{1}{3}}$, the canonical {\rm
2}-form $U$ can be obtained as:
\begin{equation} \label{determ-U-real}
U = U[W] \equiv \frac{1}{\chi \sqrt{\chi + f}}\left((\chi + f) \, F
+ \tilde{f} \, *F\right) \, ; \qquad F \equiv P(Z) \, ,
\end{equation}
where $Z$ is an arbitrary 2-form and
\begin{equation}
\hspace{-1cm} P \equiv W - \rho \, G \, , \quad \chi \equiv
\sqrt{f^2 + \tilde{f}^2} \, , \quad f \equiv \tr F^2 \, , \quad
\tilde{f} \equiv \tr (F \! \cdot \! *F) \, . \label{determ-U-real-2}
\end{equation}
\end{lemma}

\end{document}